# AUTOREGRESSIVE MODELING OF CODING SEQUENCE LENGTHS IN BACTERIAL GENOME

VASILE V. MORARIU [1,2], LUIZA BUIMAGA-IARINCA [1]

[1] *National Institute of Research and Development for Isotopic and Molecular Technologies, Department of Molecular and Biomolecular Physics, 400293, Cluj-Napoca, Romania, vvm@itim-cj.ro*

[2] *Romanian Academy of Scientists, 54, Splaiul Independentei, Sector 5, 050094 Bucharest, Romania*

Previous investigation of coding sequence lengths (CDS) in the bacterial circular chromosome revealed short range correlation in the series of these data. We have further analyzed the averaged periodograms of these series and we found that the organization of CDS can be well described by first order autoregressive processes. This involves interaction between the neighboring terms. The autoregressive analysis may have great potential in modeling various physical and biological processes like light emission of galaxies, protein organization, cell flickering, cognitive processes and perhaps others.

## 1. Introduction

Bacterial chromosome has a relatively simple structure consisting of a succession of coding (CDS) and non-coding sequences. Each sequence consists of a specific series of types and number of bases. The coding sequences represent genes and they dominate the total content of bases in the bacterial chromosome. These genes, in turn, correspond to various proteins. Recent work revealed the important significance of the length and distribution of proteins (which is similar to the coding sequences) [1-3]. This is due to the fact that there is a profound relationship between protein length distributions and the mechanism of protein length evolution. As a result the protein length distribution (or CDS lengths distribution) represents a comprehensive record of the evolutionary history of a species. The genome size of any species can be calculated only if its protein length distribution is known. Furthermore it is possible to calculate the non-coding size according to data of coding DNA. In the case of prokaryotes, the genome sizes, which are proportional to the gene numbers, can indicate the complexity of prokaryotes because the non-coding DNA contents are about the same for prokaryotes. Thus, the larger genome size is, the more complexity of prokaryotes is [3].

The earlier studies assumed that the organization of the genome is random i.e. the succession of the CDS lengths consist of uncorrelated data. However later investigations suggest that some order exists in the CDS length series [4-7].

The Detrended Fluctuation Analysis (DFA) of CDS length series suggests that bacterial genomes are short range correlated [7]. However when the same series are subjected to Fourier Transform they produce a spectrum which appears to be linear in double log plots. This apparent scale invariance means that the series are long range correlated data which contradicts DFA results. One aim of this study was to solve this apparent contradiction. The solution of the problem lays mainly in Mandelbrot's general suggestion that a spectrum should be averaged out first in order to uncover its meaning [8-9]. Indeed the averaging procedure of the spectrum revealed its nonlinear characteristic. Further we re-

*Autoregressive modeling of coding sequence lengths in bacterial genome*

port here that such spectra are described by autoregressive models which represent short range correlation.

## 2. Bacterial genome data and strategy of analysis

Each species of bacteria or archaea is characterized by a number of *n* coding sequences each having a specific length $l_k$ where $k=1...n$. The length is expressed in number of bases. Each DNA molecule is known to consist of two strands *plus* and *minus* or *leading* and *lagging strands*. The following series of CDS lengths are considered in our analysis: A) The full series consisting of the natural sequence of lengths as they succeed in the genome, B) The series of CDS lengths in the strands *plus* and *minus* respectively. Therefore the subject of our analysis are the series *l(+/-)*, *l(+)* and *l(-)*. The data were extracted from the web site of European Molecular Biology Laboratory (EMBL) by using simple programs. An example of series for a bacteria species is illustrated in fig.1.

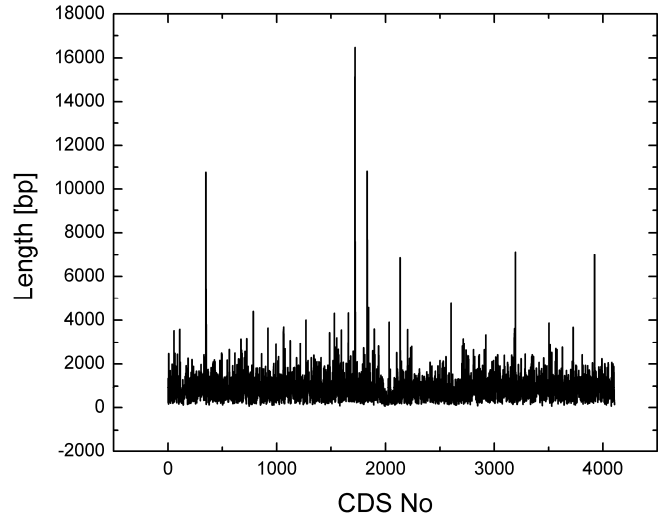

Figure 1 Coding sequence length series for *Bacillus* subtilis. The series refer to the full genome series denoted as l(+/-).

These series were analyzed by DFA and by an autoregressive model. The analysis resulted in two parameters: alpha – the correlation exponent and phi – the interaction factor respectively (see below).

It was considered of interest to explore comparatively the genome organization of some bacteria which are often used in explorative studies. We have also considered several other species of bacteria in order to see if differences exist among them. Some of bacteria exist as different cell strains so that various cell strains of the same species were analyzed. Although archaea forms a distinct domain they have a similar structure compared to the domain of bacteria so we included in our comparative analysis an archaea species. The full list of analyzed species is as follows: *Archaeoglobus fulgidus, Bacillus subtilis, Bacillus cereus ATCC14, Bacillus halonduras, Escherichia coli O157H7 EDL933, E.coli O157:H7 str. Sakai, E.coli str.K12 substr.MG1655 W3110, E.coli UTI 89, E.coli str. K-12 W311, E.coli APEC O1, Haemophilus influenzae 86, H. influenzae ATC, H. influenzae*

*PittEE, H.influenzae PittGG, Helicobacter pylori 26695, Helicobacter pylori HPAG1, Helicobacter pylori J99.*

First we present the result of the DFA analysis which proves that the series are short range correlated. Then we calculate the periodograms of the series and analyze them by an autoregressive model. Finally we compare the outcome of these types of analyses. While DFA simply shows that there is a correlation order in the bacterial genomes, the use of the autoregressive model is to show how the order is achieved in the genomes. This reveals that the length of CDS terms interact and the strength of their interaction. The succession of operations can be summarized as following: *i*) Detrend the series of data by subtracting various degrees of polynomial fits; *ii*) Discrete Fourier transform of the series; *iii*) Periodogram averaging using 1-21 terms; *iv*) Fit the spectrum to an AR(1) model. The resulting parameters are the interaction factor $\varphi$ and the dispersion $\sigma$. Their values depend on the degree of the polynomial fit used for the detrending procedure, and on the number of spectral terms used for averaging procedure. If the data can be described by an AR(1) model, choose the final values of $\varphi$ and $\sigma$, by analyzing the plot of $\varphi$ and $\sigma$ against the polynomial degree and the number of averaging terms. This procedure was applied to *l(+-)*, *l(+)* and *l(-)* series

## 3. Methods of analysis

### 3.1. *Detrended fluctuation analysis of bacterial genomes*

Our initial method of investigation was the well known DFA. The DFA method was originally developed to investigate long-range correlation in non-stationary series [9]. First DFA integrates the series which is further divided into *n* boxes of equal length. A least square line is fit to the data in each box *n* which represents the trend in that box. The integrated time series *y(k)* is detrended by subtracting the local trend $y_n(k)$ in each box. Then the root-mean square of the resulting series is calculated as a fluctuation function:

$$F(n) = \sqrt{\frac{1}{N}\sum_{1}^{N}[y(k) - y_n(k)]^2}$$

Here *N* is the number of terms. *F(n)* typically increases with box size *n* and a linear relationship on a double log graph indicates the presence of scaling:

$$F(n) \propto n^{\alpha}$$

The value of α exponent is 0.5 for uncorrelated data, then values ranging between $0.5 < \alpha < 1$ indicates a long-range correlation while values between $0 < \alpha < 0.5$ represents anti-correlation. The key step in a long-range correlation decision is that DFA plot should be linear over the whole range of box sizes covering the series. Our earlier investigation was prompted by the observation that the DFA plot of CDS length series was generally non-linear over the whole chromosome series [7]. Strictly speaking a DFA plot should be linear if long-range correlation would extend over the whole chromosome of over, at least, two orders of magnitude of the box size *n*. Nonlinearity of the DFA plot can be a simple and direct evidence for short-range correlation as recently shown by a DFA inves-

*Autoregressive modeling of coding sequence lengths in bacterial genome*

tigation of autoregressive process [10]. An example of a DFA plot for a *l(+)* series is illustrated in figure 2. The plot is clearly nonlinear.

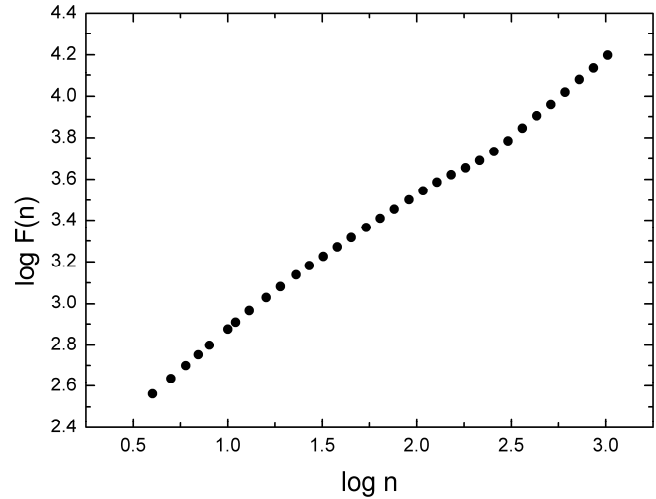

Figure 2 Detrended fluctuation analysis for the plus strand of Bacillus subtilis CDS length series.

A further simple way to check for short-range organization of the series is to apply a segmentation procedure of the CDS series and analyze the correlation in these segmented series by DFA. The correlation exponents should be unchanged if long-range correlation was operative. Short range correlation would result in different correlation exponents for different segments [10].

### 3.2. *Autoregressive modeling of the correlation spectra*

There is a key problem which recommends the analysis of the data by correlation spectroscopy. A simple Fourier transform results in noisy spectrum which can be fitted by a straight line. This is equivalent to scale invariance or a fractal like structure. However averaging the spectrum (as strongly recommended by Mandelbrot) results in non linear spectrum which contradicts the fractal like interpretation. This is illustrated in figure 4 for a bacteria species. Such a non linear spectrum is characteristic for autoregressive processes which can be easily described in an analytical way as shown below.

A discrete stochastic process $\{X_n, n=0, \pm 1, \pm 2,...\}$ is called autoregressive process of order *p*, denoted AR(*p*), if $\{X_n\}$ is stationary and for any *n*:

$$X_n - \varphi_1 X_{n-1} - ... - \varphi_p X_{n-p} = Z_n \qquad (1)$$

where $\{Z_n\}$ is a Gaussian white noise with zero mean and variance $\sigma^2$. The real parameters $\varphi_i$, $i=1,..,p$, can be interpreted as a measure of the influence of a stochastic process term on the next term after *i* time steps. The properties of AR(*p*) processes have been studied in detail and they are the basis of the linear stochastic theory of time series [11] and [12]. Equation 1 has a unique solution if the polynomial $\Phi(z)=1-\varphi_1 z-...-\varphi_p z^p$ has no roots *z* with $|z|=1$. If in addition $\Phi(z) \neq 1$ for all $|z| > 1$, then the process is causal, i.e. the random variable $X_n$ can be expressed only in terms of noise values at previous moments and at the same moment.
The spectral density of an AR($p$) process is:

$$f(\upsilon) = \frac{\sigma^2}{2\pi} \frac{1}{|\Phi(e^{-2\pi i \upsilon})|^2}, -0.5 < \upsilon \leq 0.5, \qquad (2)$$

where $\upsilon$ is the frequency. For an AR(1) process, the spectral density in equation 2 becomes:

$$f(\upsilon) = \frac{\sigma^2}{2\pi} \frac{1}{1 + \varphi^2 - 2\varphi \cos 2\pi \upsilon}, -0.5 < \upsilon \leq 0.5, \qquad (3)$$

where $\varphi$ is the only parameter $\varphi_i$ in this case. The above mentioned formulas are valid for ideal stochastic processes of finite length.
The time series found in practice have a finite length and usually they are considered realizations of a finite sample of an AR(1) process of infinite length. Therefore, the changes of the equations 2 and 3 have to be analyzed for a sample with finite length $\{X_n, n=1, 2, ..., N\}$ extracted from an infinite process $\{X_n, n=0, \pm1, \pm2, ...\}$. A detailed analysis of the power spectrum of the AR(1) process and the influence of the finite length is contained in [13]. In this paper some of the main conclusions are discussed.
The sample estimator of the spectral density is the periodogram:

$$I_N(\upsilon) = |A_N(\upsilon)|^2, \qquad (4)$$

where $A_N(\upsilon)$ is the discrete Fourier transform of the sample:

$$A_N(\upsilon) = \frac{1}{\sqrt{N}} \sum_{n-1}^{N} X_n e^{2\pi i n \upsilon}, \qquad (5)$$

Since the sample contains a finite number of components, there are only $N$ independent values of $A_N(\upsilon)$ and $I_N(\upsilon)$. Usually, these values are computed for the Fourier frequencies $\upsilon_j = j/N$, where $j$ is a integer satisfying the condition $-0.5 < \upsilon_j \leq 0.5$. The periodogram of an AR($p$) process is an unbiased estimator of the spectral density:

$$\lim_{N \to \infty} \langle I_N(\upsilon_j) \rangle = 2\pi f(\upsilon), \qquad (6)$$

where $(\upsilon_j - 0.5N) < \upsilon \leq (\upsilon_j + 0.5N)$ [13]. Hence, increasing the sample length $N$, while the time step t is kept constant, the average periodogram becomes a better approximation of the spectral density (equation 6). However, a single periodogram is not a consistent estimator, because it does not converge in probability to the spectral density, i.e. the standard deviation of $I_N(\upsilon_j)$ does not converge to zero, and two distinct values of the periodogram are uncorrelated, no matter how close the frequencies are when they are computed.
Usually, the spectral density and the periodogram are plotted on a log-log scale. The logarithmic coordinates strongly distort the shape of the graphic because by applying the logarithm, any neighborhood of the origin is transformed into an infinite length interval and the value of *f*(0) cannot be plotted. For a sample with *N* terms, the first value of the spectral density is obtained for the minimum frequency $v_{min} = 1/N$. Figure 2a shows the spectral density in equation 3 for $N = 1024$, $\sigma = 1$ and different values of the parameter $\varphi$. One can observe that the AR(1) processes for higher φ values can be easily approximated by a fractal-like behavior. For $\varphi = 0.90$ and especially for $\varphi = 0.99$, a significant part of the power spectrum is almost linear with a slope equal to −2, which corresponds to $\beta = 2$. A significant part of the spectrum could be regarded as linear for smaller value of $\varphi$ (for example $\varphi = 0.5$ in figure 3).

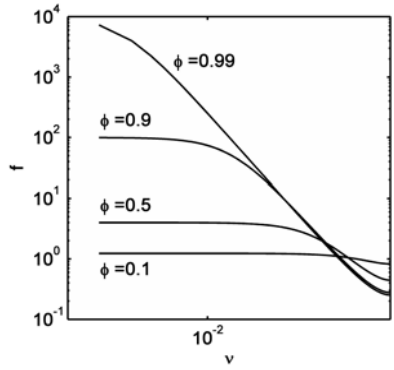

Figure 3 The spectral density of an AR(1) process for N = 1024, σ =1 and different values of the interaction factor among close terms φ.

For small frequencies, the AR(1) spectral density is strongly stretched in log-log coordinates such that a plateau appears (figure 2a) with a value given by:

$$f(0) = \frac{\sigma^2}{2\pi(1-\varphi)^2} \qquad (8)$$

From equation 3 it follows that the plateau corresponds to the small values of *v*, when the variable term at the denominator can be neglected in comparison with the constant term. Using the quadratic approximation of the cosine function, the condition that the graph of the AR(1) power spectrum has a plateau $v < (1 - \varphi)/2\pi\sqrt{\varphi}$ is obtained. If $\varphi$ tends to 1, the plateau appears at smaller values of the frequency. Therefore, if *N* is large enough, the periodogram of an AR(1) sample has a plateau at small frequencies (if *N* is large, then $v_{min} \to 0$).

A time series $\{x_n, n=1, 2, ..., N\}$ as a realization of a sample $\{X_n, n=1, 2, ..., N\}$ from an AR(1) process is considered. Applying the discrete Fourier transform (equation 4) to the time series $\{x_n\}$, and then computing the periodogram (equation 5), values randomly distributed around the spectral density (equation 3) of the AR(1) are obtained. Since the periodogram is not a consistent estimator, by increasing the length *N* of the sample, the periodogram fluctuations around the theoretical spectral density are not reduced. Consistent estimation of the spectral density may be obtained using averaging of the periodogram on intervals with length of magnitude order of $\sqrt{N}$ [13]. Choosing the optimum weight function is a difficult task, because, if the periodogram is smoothed too much, then the bias with respect to the theoretical spectrum can become large. From various weight functions [16] the simplest one is used, i.e. the averaging with equal weights on symmetric intervals containing $M$ Fourier frequencies, with $M = 1, 3, 5, ..., 21$. Then, the averaged periodogram contains $N − M +1$ values, because for the first and last $(M − 1)/2$ values of the periodogram, the symmetric averaging cannot be performed.

Let us consider that an AR(1) model for an averaged periodogram is to be found, i.e. to find the values of the parameters $\varphi$ and $\sigma$. The minimum of the quadratic norm of the difference between the averaged periodogram and the theoretical spectral density of the AR(1) model has to be determined. The sample standard deviation of the time series and $\varphi = 0$ are used as initial values for the optimization algorithm. All calculations were performed by using MATLAB [15].

Many series of data have non-stationary characteristics, so the application of Fourier transform to the data results in misleading spectra. A common procedure to avoid this complication is to use detrended fluctuation analysis[9]. This results in a correlation exponent free of the correlation introduced by the trend. However, in our case it is essential to obtain the corresponding spectrum, as the shape of the spectrum gives the relevant information (if either a power-law or a non-power law is operative). Consequently, an important preliminary step is to remove non-stationary characteristics in the series. We performed detrending by subtracting a polynomial fit from the original series. The problem is to determine the right polynomial fit. We performed 1 to 20 degree polynomial fits and generally found that a polynomial degree around 10 gives the most reliable result for $\varphi$ and $\sigma$. The values of $\varphi$ and $\sigma$ also depended on the averaging procedure of the spectra so that optimizing the values of $\varphi$ and $\sigma$ involved optimizing both the detrending and the averaging procedures. The above mentioned polynomial fitting was chosen as its accuracy is comparable to the moving average method and to an automatic method for the estimation of a monotone trend. The same work also showed that polynomial fitting for a $1/f$ noise proved to have the best performance [14].

## 4.  Results and Discussion

Examples of non-averaged and averaged periodogram for the CDS length series *for Bacillus subtilis* species are illustrated in figure 4. The non-averaged periodogram (figure 4a) can be well fitted by a straight line which can mistakenly interpreted as a long range correlation. However the real shape of the periodogram is revealed after averaging. (figure 4 b). The averaged periodogram is well fitted by an AR(1) process.

The corresponding AR(1) parameters φ as well as the correlation exponents resulting from DFA analysis are included in table 1 and 2 for a range o bacteria an archaea species. The parameter φ may vary between 0<φ<1. In the case of bacteria and archaea presented in table 1 this parameter varies between 0.52 for *Bacillus subtilis* and almost 0 for some strands of *Haemophilus influenzae* and *Helicobacter pilori*. This means that the strenght of interaction between the the length of CDS terms may vary widely from a medium value down to almost no interaction. Further this interaction is sensitive to the *leading* or *lagging* strands and also to the species strain. On the other hand the correlation exponent (table 2) generally has a rather low value indicating a weak correlation among the terms of the series.

Higher order autoregressive processes where also tested. This involved additional interaction between terms at higher distances in the series. However it was concluded that the AR(1) modeling of the series proved to be quite satisfactory.

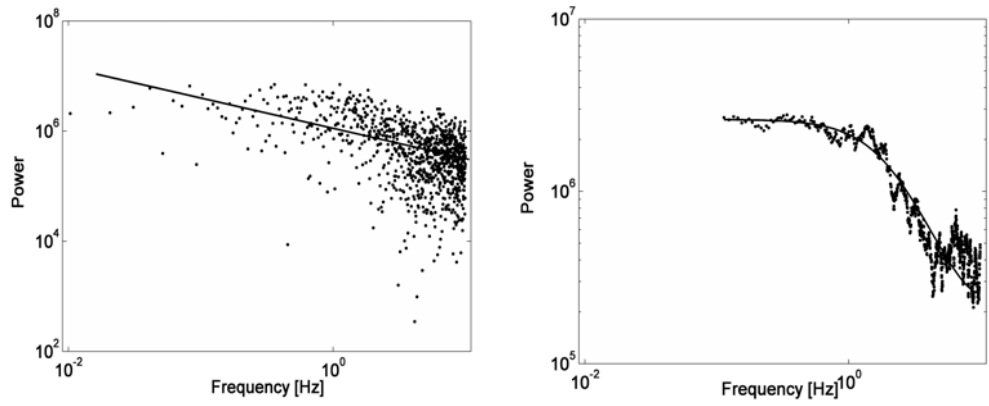

Figure 4 Periodogram for the length series of coding sequences of the plus (or leading strand) of *Bacillus subtilis*. a) Non-averaged periodogram; b) Averaged periodogram and fitting with a first order autoregressive process.

Table 1 The interaction strength φ of the first order autoregressive modeling of CDS lengths periodogram.

| Species | + strand | - strand | both strands |
|---|---|---|---|
| *Archaeglobus Fulgidus* | φ = 0.01 | φ = 0.16 | φ = 0.14 |
| *Bacilus subtilis* | φ = 0.52 | φ = 0.43 | φ = 0.44 |
| *B Cereus ATCC14* | φ = 0.19 | φ = 0.05 | φ = 0.13 |
| *B Halonduras* | φ = 0.11 | φ = 0.10 | φ = 0.16 |
| *E. Coli O157H7* | φ = 0.15 | φ = 0.21 | φ = 0.24 |
| *E. Coli Sakai* | φ = 0.21 | φ = 0.26 | φ = 0.20 |
| *E. Coli K12* | φ = 0.12 | φ = 0.20 | φ = 0.18 |
| *E. Coli UTI* | φ = 0.14 | φ = 0.21 | φ = 0.24 |
| *E Coli W311* | φ = 0.17 | φ = 0.14 | φ = 0.18 |
| *E Coli APEC* | φ = 0.18 | φ = 0.15 | φ = 0.19 |
| *H Influ 86* | φ = 0.07 | φ = 0.13 | φ = 0.13 |
| *H Influ ATC* | φ = 0.11 | φ = 0.07 | φ = 0.12 |
| *H Influ Pitt EE* | φ = 0.11 | φ = 0.05 | φ = 0.11 |
| *H Influ. Pitt GG* | φ = 0.13 | φ = 0.00 | φ = 0.10 |
| *H Pylori 266* | φ = 0.05 | φ = 0.00 | φ = 0.02 |
| *H Pylori HP* | φ = 0.05 | φ = 0.02 | φ = 0.03 |
| *H Pylori J99* | φ = 0.04 | φ = 0.03 | φ = 0.04 |

Table 2 The correlation exponent for different species of bacteria and archaea resulting from the detrended fluctuation analysis

| Species | + strand | - strand | both strands |
|---|---|---|---|
| *Archaeglobus Fulgidus* | α = 0.50 | α = 0.60 | α = 0.57 |
| *Bacilus subtilis* | α = 0.84 | α = 0.67 | α = 0.79 |
| *B Cereus ATCC14* | α = 0.60 | α = 0.50 | α = 0.55 |
| *B Halonduras* | α = 0.56 | α = 0.55 | α = 0.56 |
| *E. Coli O157H7* | α = 0.56 | α = 0.60 | α = 0.60 |
| *E. Coli Sakai* | α = 0.56 | α = 0.62 | α = 0.56 |
| *E. Coli K12* | α = 0.55 | α = 0.57 | α = 0.58 |
| *E. Coli UTI* | α = 0.56 | α = 0.57 | α = 0.60 |
| *E Coli W311* | α = 0.59 | α = 0.54 | α = 0.57 |
| *E Coli APEC* | α = 0.56 | α = 0.55 | α = 0.58 |
| *H Influ 86* | α = 0.53 | α = 0.58 | α = 0.55 |
| *H Influ ATC* | α = 0.53 | α = 0.55 | α = 0.52 |
| *H Influ Pitt EE* | α = 0.55 | α = 0.55 | α = 0.54 |
| *H Influ. Pitt GG* | α = 0.62 | α = 0.50 | α = 0.54 |
| *H Pylori 266* | α = 0.50 | α = 0.50 | α = 0.51 |
| *H Pylori HP* | α = 0.54 | α = 0.49 | α = 0.51 |
| *H Pylori J99* | α = 0.53 | α = 0.52 | α = 0.54 |

It is obvious that a relationship should exist among the correlation exponents α and the interaction factor φ for different species. This is illustrated in figure 5 for the three kind of series. It can be seen that the relationship is linear for *l(+)* and *l(-)* series while for the *l(+-)* series is non linear (figure 5). Non averaged periodograms give very similar results (not shown). This might suggests that the non-stationary contribution to the series are not important for the AR (1) modeling of the series.

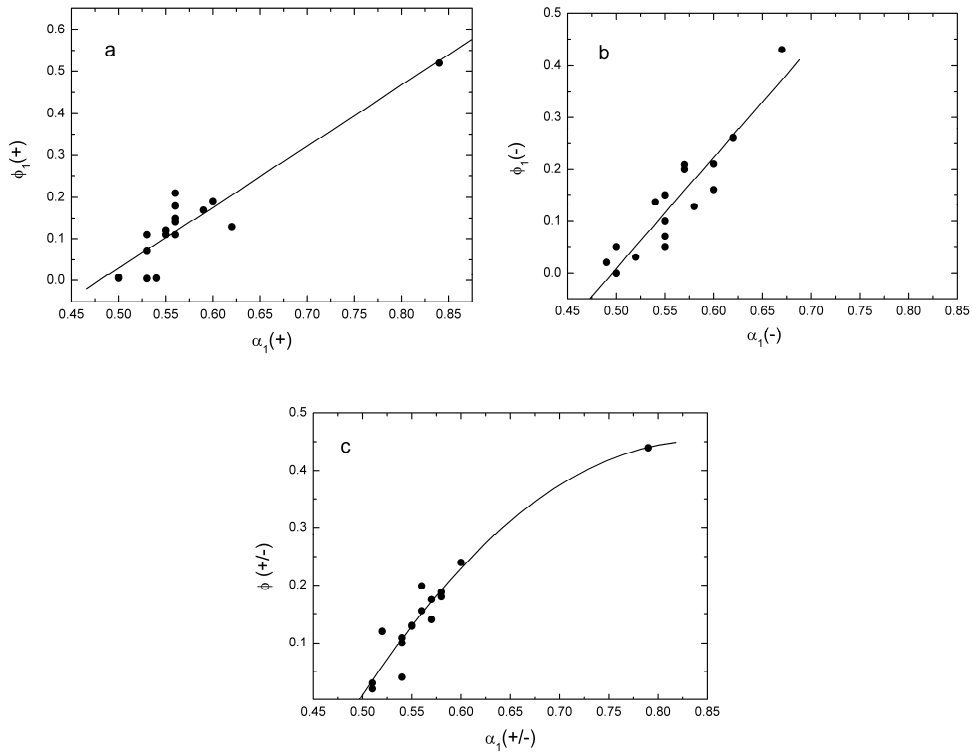

Figure 5 The relationship between the correlation exponent *alpha 1* (resulting from DFA analysis) and the strength of interaction *phi 1* (first order autoregressive modeling). a) the *strand +* data, b) *strand -* and c) *strand +/-* of *Bacillus subtilis*.

Parameter φ of the autoregressive fitting seems to be more sensitive compared to the correlation exponent. The impression at this stage is that first order autoregressive fitting of the CDS series is the best way to describe the order in these series. We further attempted to describe the order in the series of very different kind of phenomena which presented similar features of the spectrum. Such examples refer to length of protein series, biological cell flickering or cognitive phenomena. [9]. Therefore autoregressive processes proved to be a useful model for describing the order in these phenomena.

**Acknowledgements**

The work was supported by the Romanian Agency for Scientific Research. The authors are indebted to Calin Vamos for computation work.